\documentclass[fleqn,10pt]{wlscirep}
\title{Direct growth of low-doped graphene on Ge/Si(001) surfaces}

\author[1,*]{J. Dabrowski}
\author[1,+]{G. Lippert}
\author[2]{J. Avila}
\author[3]{J. Baringhaus}
\author[4]{I. Colambo}
\author[1,++]{Yu. S. Dedkov}
\author[5]{F. Herziger}
\author[1]{G. Lupina}
\author[5]{J. Maultzsch}
\author[1]{T. Schaffus}
\author[1,6]{T. Schroeder}
\author[1,6]{M. Sowinska}
\author[3]{C. Tegenkamp}
\author[4]{D. Vignaud}
\author[2]{M.-C. Asensio}
\affil[1]{IHP, Im Technologiepark 25, 15236 Frankfurt (Oder), Germany}
\affil[2]{Synchrotron SOLEIL, Saint Aubin, BP 48, 91192 Gif-sur-Yvette, France}
\affil[3]{Institut f\"ur Festk\"orperphysik, Leibniz Universit\"at, Appelstr. 2, 30167 Hannover, Germany}
\affil[4]{IEMN, Av. Poincar\'e CS 60069, 59652 Villeneuve d'Ascq Cedex, France}
\affil[5]{Institut f\"ur Festk\"orperphysik, TU Berlin, Hardenbergstr. 36, 10623 Berlin, Germany}
\affil[6]{BTU Cottbus-Senftenberg, Konrad Zuse Str. 1, 03046 Cottbus, Germany}

\affil[*]{E-mail: dabrowski@ihp-microelectronics.com}
\affil[+]{E-mail: lippert@ihp-microelectronics.com}
\affil[++]{E-mail: dedkov@ihp-microelectronics.com}
\begin{abstract}
The practical difficulties to use graphene in microelectronics and optoelectronics is that the available methods to grow graphene are not easily integrated in the mainstream technologies. A growth method that could overcome at least some of these problems is chemical vapour deposition (CVD) of graphene directly on semiconducting (Si or Ge) substrates. Here we report on the comparison of the CVD and molecular beam epitaxy (MBE) growth of graphene on the technologically relevant Ge(001)/Si(001) substrate from ethene (C$_2$H$_4$) precursor and describe the physical properties of the films as well as we discuss the surface reaction and diffusion processes that may be responsible for the observed behavior. Using nano angle resolved photoemission (nanoARPES) complemented by transport studies and Raman spectroscopy, we report the direct observation of massless Dirac particles in monolayer graphene, providing a comprehensive mapping of their low-hole doped Dirac electron bands. The micrometric graphene flakes are oriented along two predominant directions rotated by $30^\circ$ with respect to each other. The  growth mode is attributed to the mechanism when small graphene ``molecules'' nucleate on the Ge(001) surface and it is found that hydrogen plays a significant role in this process.
\end{abstract}
\begin{document}

\flushbottom
\maketitle
\thispagestyle{empty}

\section*{Introduction}

Graphene is widely supposed to be a material on the way to enable new development of many modern technologies~\cite{Geim:2007a,Geim:2009}. This includes microelectronics and optoelectronics, which require graphene of the highest quality, inexpensively produced and processed on large areas~\cite{Bae:2010,Bonaccorso:2013iy,Novoselov:2013hw}. Such graphene can be grown on Cu or on Ni~\cite{Yu:2008,Kim:2009a,Li:2009,Bae:2010}, but then it must be transferred to the target wafer. The transfer process is problematic; one of the reasons is contamination with residual metal atoms~\cite{Ambrosi:2014gl,Lupina:2015je}. Therefore, researchers examine the prospects of growing graphene on microelectronics-friendly substrates, such as silicon or germanium~\cite{Ochedowski:2012bx,ThanhTrung:2014hd,Dang:2015fk,Zeng:2013jp,Cavallo:2014fp,Wang:2013fq,Lee:2014dv,Jacobberger:2015de,Lippert:2014fc,McElhinny:2016gw,Pasternak:2016jt}. Direct growth of graphene on Si is hindered by high reactivity of Si against C~\cite{ThanhTrung:2014hd}. In contrast to Si, Ge does not form stable carbide. However, much higher growth temperatures are needed for CVD graphene on Ge than on insulating hexagonal BN, indicating that the formation of C-Ge bonds plays a major role in the former process~\cite{Yang:2013eva,Lee:2014dv}.

Graphene films grown on germanium may be used in two ways. First, if large area graphene (like $300-450$\,mm in diameter) is obtained, the film may be transferred to the target wafer at any point of the fabrication process (i.\,e., during front end of line or back end of line steps) without the hazard that the production tools become contaminated. In practice this means that the growth must be performed on Ge(001)/Si(001) wafers, because Ge wafers of this size are not commercially available and it would most probably be prohibitively expensive to develop and introduce a Ge wafer growth technology solely for this purpose. The second way to use graphene films grown on germanium may be open when graphene can be grown directly on device-size Ge(001) islands pre-deposited on the Si(001) substrate. For example, the width of the active region of a graphene base transistor may be smaller than a micron, and its whole area may be of the order of $\mu\mathrm{m}^2$~\cite{Mehr:2012cv,Vaziri:2013er}. 

Graphene growth on the technologically relevant Ge(001) and Ge(001)/Si(001) wafers has been reported for a CH$_4$-based CVD~\cite{Wang:2013fq} and for a molecular beam epitaxy (MBE) (atomic carbon)~\cite{Lippert:2014fc}, respectively. In both cases, the structural and electrical quality of graphene significantly improves at temperatures so high that the Ge substrate begins to melt. Besides causing practical problems associated with strong surface roughening, that takes place at such temperatures, this again highlights the importance of the C-Ge interaction. In the present manuscript, we report on the observation that illustrates the influence of this interaction on the growth process -- the appearance of two orientational domains. These two domains are revealed by the presence of two sets of Dirac cones in the electronic structure of the graphene film as deduced from the nanoARPES data. The interaction of the C$_2$H$_4$ precursor with the Ge(001) surface is then analyzed in the framework of \textit{ab initio} density functional theory (DFT) calculations and conclusions for the microscopic mechanisms of the growth process are drawn. These mechanisms are then compared to those expected for the growth of graphene on Ge(001) from atomic carbon.

\section*{Results}

\begin{figure}
\includegraphics[width=\textwidth]{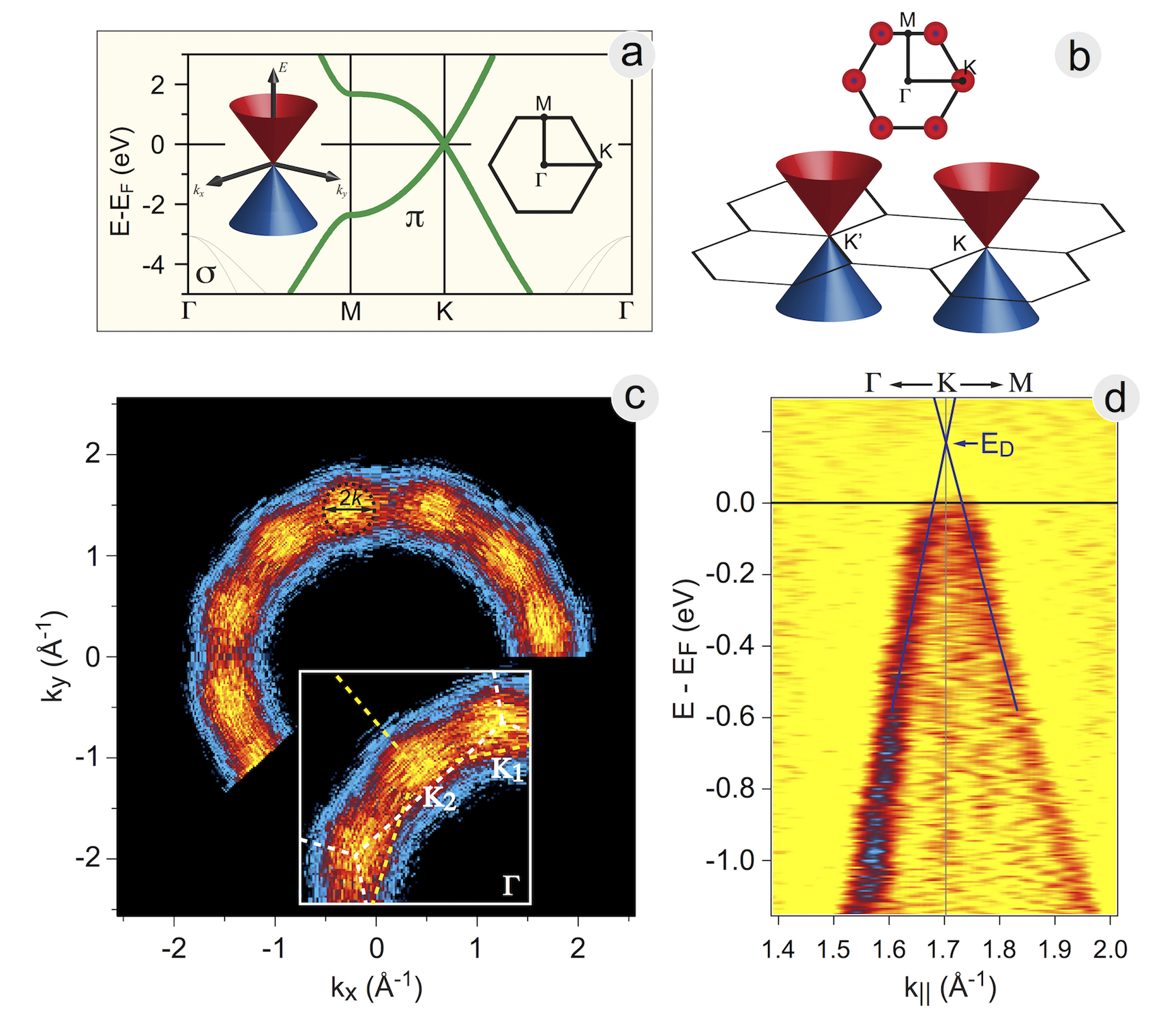}\\
\caption{\label{ARPES} \textbf{Two domain orientations in CVD graphene on Ge(001) as measured in nano-ARPES experiments.}  (a) Calculated band structure of graphene around $E_F$. (b) The Brillouin zone shape and the location of the Dirac cones. There are six $\mathrm{K}$ points on the zone boundary, one in each of the hexagon corners: the cones are distributed in a ring around the $\Gamma$ point and separated by $60^\circ$. (c) ARPES map of density of states at $E_F$ ($h\nu=100$\,eV, linear polarization), revealing a ring of Dirac cones with $30^\circ$ separation, meaning that within the X-ray spot there are two orientational domains of graphene, each occupying approximately the same area. Inset shows the zoom in the reciprocal space for several spots around two $\mathrm{K}$ points originated from two graphene flakes. (d) Second derivative of the ARPES intensity map ($h\nu=30$\,eV, circular polarization) measured along one of the $\Gamma-\mathrm{K}$ directions.} 
\end{figure}

\textbf{Experiment.} The electronic structure in the vicinity of the Fermi level ($E_F$) of the CVD-produced graphene flakes on Ge(001)/Si(001) substrate was studied in the micro and nanoARPES experiments with $120$\,$\mu$m and $100$\,nm spot size resolution, respectively~\cite{Avila:2013a,Avila:2013b,Avila:2014a} (Fig.~\ref{ARPES}). As it is well-known for a free-standing graphene, the charge neutrality point (Dirac point, $E_D$) coincides with $E_F$. In the vicinity of $E_D$, the energy of graphene electronic states depends linearly on the wave-vector $k$ (Fig.~\ref{ARPES}a). These energy bands form the so-called Dirac cones at the six equivalent $\mathrm{K}$ points in the corners of the hexagonal Brillouin zone (BZ) of graphene (Fig.~\ref{ARPES}b). In case of the $n$- or $p$-doped graphene, the Fermi level is located either above or below the Dirac point, respectively. In-plane long-range disorder in the graphene lattice blurs the energy states imaged by taking the constant energy cuts in the 3D ARPES data, $I(E_B,k_x,k_y)$. Here, $E_B$ is the binding energy and $k_x$, $k_y$ are two in-plane components of the wave vector of electron. When a neutral graphene has two domains rotated with respect to each other by $30^\circ$ and the same area of each of them is illuminated by the photon beam, the Fermi surface in the ARPES experiment consists of 12 equivalent spots centred at the 12 equivalent $\mathrm{K}$-points around the BZ centre, $\Gamma$. When a neutral graphene is fully disordered or polycrystalline, the Fermi surface forms a circle centred at $\Gamma$ and similar to the one observed for highly oriented pyrolytic graphite (HOPG).

Figure~\ref{ARPES}c shows a constant energy cut taken at the Fermi level and extracted from the 3D photoemission intensity data set, $I(E,\theta,\phi)$ measured for graphene/Ge(001)/Si(001) at $h\nu = 100$\,eV (spot size $120$\,$\mu$m) as a function of two emission angles, polar $\theta$ (detector angle) and azimuth $\phi$ (in-plane sample rotation), and then converted into $I(E_B,k_x,k_y)$. One can clearly see that the photoemission at the Fermi level has a circle-like shape centred at $\Gamma$ and having equidistant intensity maxima. The radius of the circle in the reciprocal space is $1.7\mathrm{\AA}^{-1}$ and the angular distance between the maxima is $30^\circ$. This indicates that the observed photoemission intensity originates from graphene flakes with orientation distribution that has probability maxima in the directions corresponding to graphene hexagons being parallel and/or perpendicular to the surface dimer rows of the Ge(001) surface (see discussion below and Fig.~\ref{structure}).

A detailed map of photoemission intensity measured with circularly polarized light of $h\nu = 30$\,eV along the $\Gamma-\mathrm{K}$ direction for one of the preferential domains is given in Fig.~\ref{ARPES}d. Due to the fact that circularly polarized light is used, the photoemission bands along both $\mathrm{K}-\Gamma$ and $\mathrm{K}-\mathrm{M}$ directions are detected. One can also see that the photoemission intensity between two branches does not vanish; this is due to the partially random orientation of graphene flakes. However, a clear Dirac cone is visible; this is valid for all intensity spots in Fig.~\ref{ARPES}c.

These data allow us to estimate the energy distance between the Fermi level and the Dirac point, and hence the doping level of graphene: $E_F$ is located by $0.185$\,eV below $E_D$, which corresponds to the hole concentration of $2.3\times10^{12}\mathrm{cm}^{-2}$. The linear fit of the Dirac cone in the ARPES intensity map in Fig.~\ref{ARPES}d gives the Fermi velocity of $(1.36\pm0.3)\times10^6$\,m/s. This can be compared with the graphene/Ge/SiC(0001) system, where $p$-doping of graphene was obtained via intercalation of Ge atoms in graphene/SiC at $720^\circ$\,C and the hole doping of $4.1\times10^{12}\mathrm{cm}^{-2}$ was measured~\cite{Emtsev:2011fo}. The measured radius of the Fermi surface around $\mathrm{K}$ is $0.027\mathrm{\AA}^{-1}$.

\begin{figure}
\includegraphics[width=\textwidth]{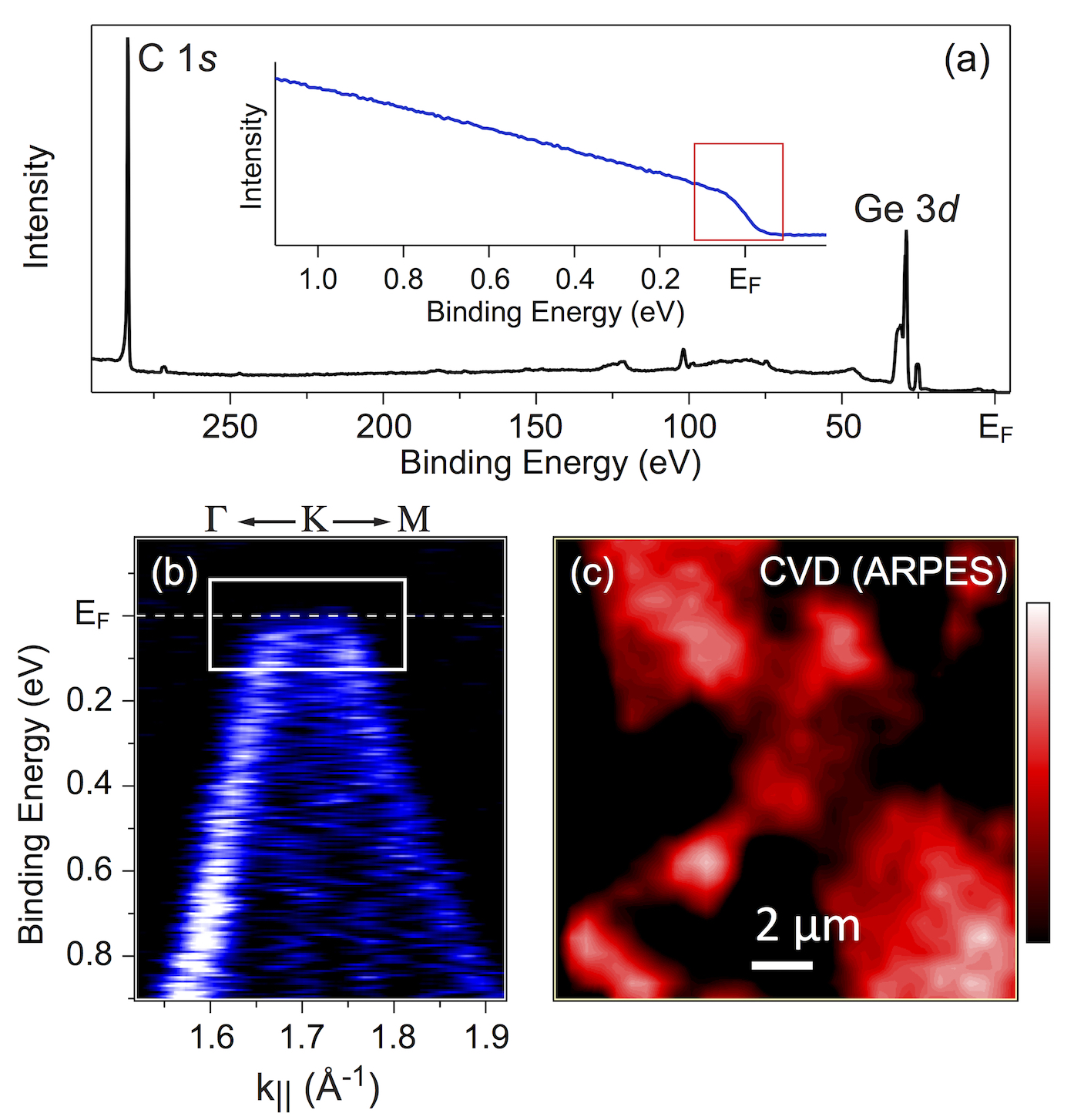}\\
\caption{\label{ARPES_maps} \textbf{Photoelectron spectroscopy (PES) characterization of grpahene on Ge(001).} (a) Survey PES spectrum of the CVD grown graphene on Ge(001)/Si(001) measured with photon energy of $h\nu=350$\,eV. Inset shows the photoemission intensity energy profile around $E_F$ obtained by integration of intensity around one of the Dirac cones of graphene (see Fig.~\ref{ARPES}d). (b) Zoom for Fig.~\ref{ARPES}d with the energy and $k$-vector window (marked by white rectangular) used in space resolved measurements. (c) Space distribution of the photoemission intensity obtained via $xy$-scan of the graphene/Ge(001) sample using the window marked in inset of (a) and (b).} 
\end{figure}

Graphene flakes distribution was probed in the real-space-resolved nanoARPES experiments. Fig.~\ref{ARPES_maps}(a) shows the respective PES survey spectrum for our sample before ARPES mapping where C\,$1s$ and Ge\,$3d$ core levels are marked. The photoemission intensity at the $\mathrm{K}$-point for one of the preferential graphene domains and in the vicinity of the Fermi level (the energy and wave-vector windows are marked as a rectangular in inset of Fig.~\ref{ARPES_maps}a and in Fig.~\ref{ARPES_maps}b) was acquired as a function of space coordinate (see Fig.~\ref{ARPES_maps}c). The photoemission intensity map indicates that the graphene film consists of flakes with size in the $\mu$m range. Indeed, such flakes can be also recognized in scanning electron microscopy (SEM) images (Fig.~\ref{SEM_maps}). The graphene flakes in the CVD and MBE prepared films appear similar in SEM, at least in the growth stage depicted in the figure, but one interesting difference between them is that the MBE prepared flakes have long straight edges, while the CVD synthesized flakes are of much more rounded shape. The flakes are separated by a material that is far less conducting. In the SEM images these areas appear as bright spots, and the ARPES spatial maps show that the density of states at the $\mathrm{K}$-point around $E_F$ in the CVD sample vanishes there. 

\begin{figure}
\includegraphics[width=0.8\textwidth]{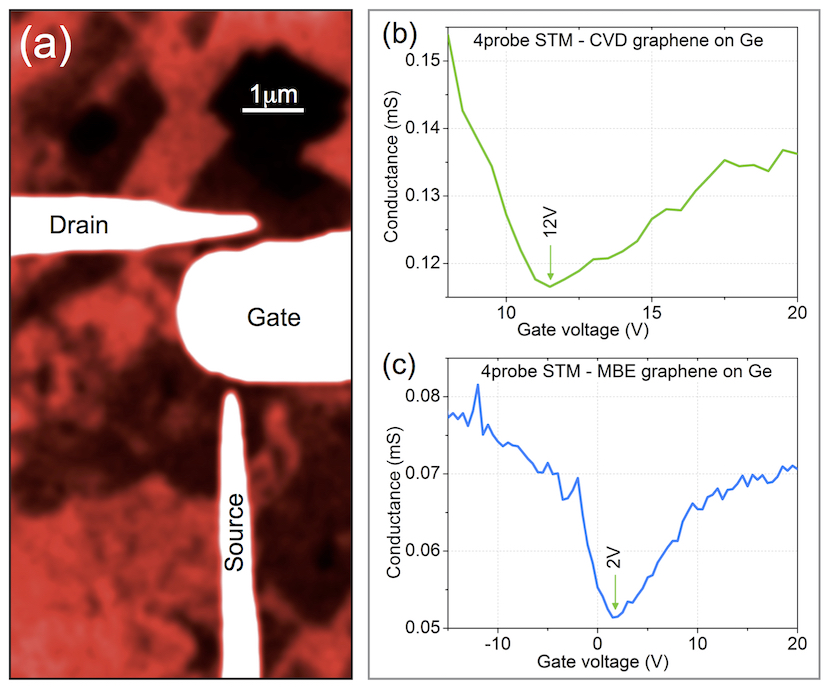}\\
\caption{\label{STM_FET} \textbf{Electrical measurements with 4-tip STM.} (a) SEM image of the tip arrangement in the FET configuration. One tip serves as a transistor gate and two other act as the source and drain contacts. (b) Graphene channel conductance as a function of the gate voltage: a CVD sample, tip-sample distance $d=15$\,nm. (c) The same for an MBE sample, $d=25$\,nm; the graphene is here separated from Ge(001) by a graphene-like interfacial layer~\cite{Lippert:2014fc}.} 
\end{figure}

Results of a electrical measurements in the field-effect transistor (FET) configuration in 4-point STM (Fig.~\ref{STM_FET}a) are consistent with the observation of $p$-type doping detected in nanoARPES experiments (cf. Fig.~\ref{ARPES}d): the Fermi level crosses the Dirac point (and the graphene channel conductance reaches its minimum) when the gate is positively biased. This direction of the shift corresponds to $p$-doped channel. The same sign of doping was detected for CVD as well as for MBE prepared samples (Fig.~\ref{STM_FET}b,c). The gate bias $V_{FB}$ at which the channel conductance reaches its minimum is a measure of the Fermi level position (with respect to the Dirac point), which in turn is determined by the doping level; therefore $V_{FB}$ measures the doping level. Comparison of the $V_{FB}$ value for CVD and MBE prepared films indicates that former films are more strongly $p$-doped. The measurements in the FET configuration allow to determine the electron mobility of the graphene channel. For this purpose, one assumes that the film area controlled by the gate tip corresponds to a square and one measures the sample-tip distance $d$ by approaching the tip towards the sample until mechanical contact is observed. The measured mobilities were lower than for high-quality graphene and amounted to $1200\pm400\,\mathrm{cm}^2/\mathrm{V}\cdot\mathrm{s}$ and $600\pm300\,\mathrm{cm}^2/\mathrm{V}\cdot\mathrm{s}$ for CVD and MBE prepared graphene/Ge(001), respectively. 

\begin{figure}
\includegraphics[width=\textwidth]{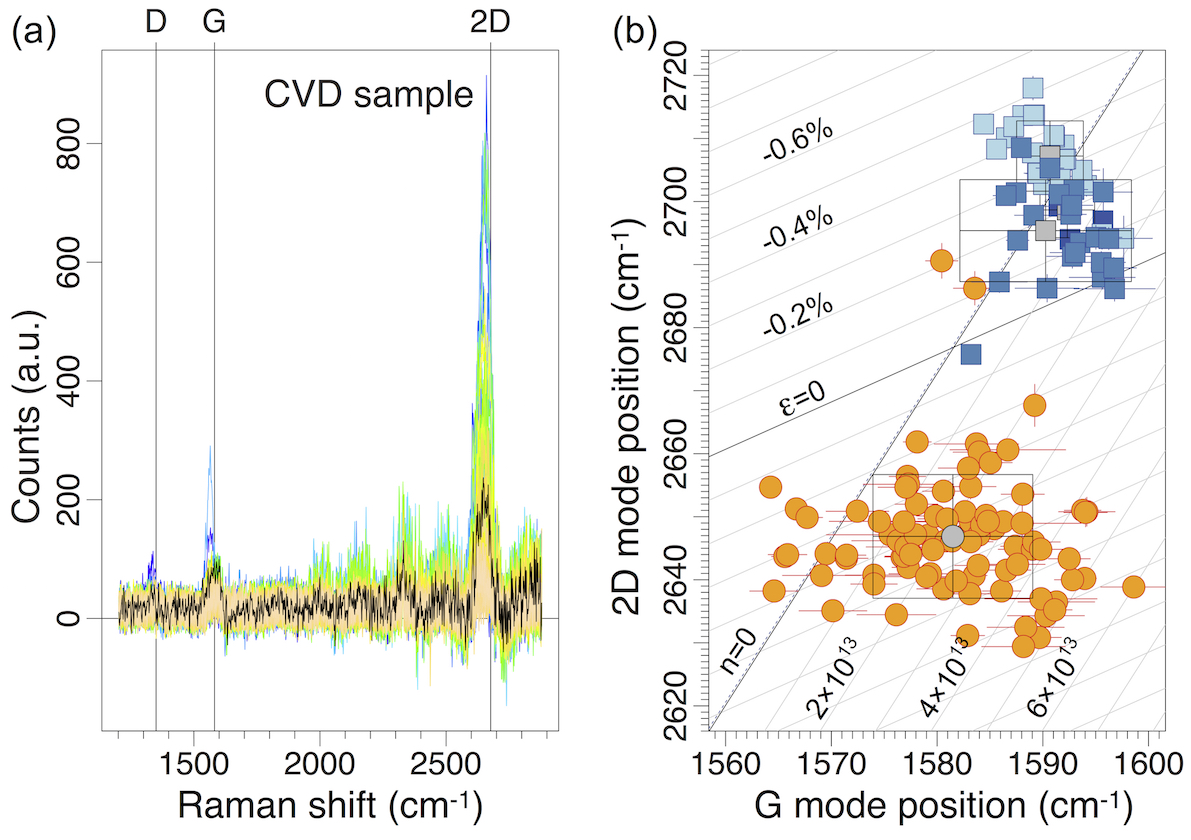}\\
\caption{\label{raman} \textbf{Raman characterization of graphene films.} (a) Collection of 130 spectra from a CVD film after background subtraction; the spectrum closest to the average is shown in black. Energies of D, G, and 2D modes in an unstrained and undoped graphene are indicated. (b) 2D-position vs. G-position plot illustrating the difference in the doping and strain levels in MBE (blue squares) and CVD (orange circles) graphene films on Ge(001). The black straight lines indicate the 2D (G) mode dependence for biaxialyl strained undoped and unstrained $p$-doped, free-standing graphene; the distances between the grey lines amount to 0.1\% and to $1\times10^{13}\mathrm{cm}^{-2}$, respectively. The average strain is compressive in the MBE film and tensile in the CVD film. The MBE graphene film is separated from Ge(001) by a graphene-like interfacial layer~\cite{Lippert:2014fc}.} 
\end{figure}

Figure~\ref{raman} summarizes the results of our Raman analysis of the MBE- and CVD-grown graphene layers on Ge(001) described above. In both cases, the phonon frequencies of the 2D and G modes differ from their reference values measured on undoped and unstrained graphene films. Following the analysis of Lee \textit{et al.}~\cite{Lee:2012gy}, we can disentangle the effects of strain and doping in our graphene films on Ge(001) (see Fig.~\ref{raman}b). Since the 2D-mode position for the MBE-grown graphene is upshifted compared to the reference value, the strain in the MBE-grown graphene is compressive, assuming that the strain is biaxial; the average strain can be estimated to approximately -0.3\%. In contrast, the strain in the CVD-grown graphene films is tensile and can be evaluated to about 0.4\%. If we would instead assume uniaxial strain for both samples, the strain would amount to -0.8\% and 1.4\% for the MBE- and CVD-grown graphene, respectively. However, such large uniaxial strain would lead to a splitting of the double-resonant 2D mode by more than $20\,\mathrm{cm}^{-1}$~\cite{Frank:2011cu}. This is in contrast to our experimental data, where a symmetric, Lorentzian-shaped 2D mode is observed during all measurements. Thus, strain in our samples is likely to be biaxial.

In accordance with the electrical 4-point STM measurements, the doping level of the CVD graphene film is higher than that of the MBE-grown film as deduced from the Raman spectra measurements. A rough estimate yields the average doping below $1\times10^{12}\mathrm{cm}^{-2}$ for the MBE graphene and around $3\times10^{13}\mathrm{cm}^{-2}$ for the CVD sample. The presence of data points with G-mode positions on the left side of the zero-doping line can be partially explained with the anomalous phonon softening of the G mode for Fermi-level shifts close to the Dirac point~\cite{Lee:2012gy,Pisana:2007cl,Lazzeri:2006kz}. Additionally, we have to consider other effects than strain and doping to affect the 2D- and G-mode positions. In particular, the background subtraction, which is necessary due to strong luminescence from the substrate, introduces an additional uncertainty to the strain/doping analysis that is in the order of 0.2\% and $5\times10^{12}\,\mathrm{cm}^{-2}$.

\textbf{Theory.} Although carbon and germanium form no stable carbide phase (GeC), the chemical interaction between the Ge(001) surface and C atoms is strong. This is particularly pronounced when atomic carbon is deposited~\cite{Lippert:2014fc}. In this case, there is a significant probability that a mobile C atom becomes immobilized by substituting a Ge atom in a Ge surface dimer (Ge$_2$), whereby a mobile Ge adatom is released. The process is reversible: a Ge adatom may eject the C atom from a C-Ge surface dimer. The mobile and highly reactive C interstitially produced in this process can diffuse under the surface. The mobile species controlling the MBE growth mode of graphene on Ge(001) are C atoms, short chains consisting of few C atoms, and Ge adatoms; the latter are attracted to the edge of growing graphene. Carbon dimers (C$_2$) are trapped by Ge dimer vacancies (DV) to become graphene seeds; a similar role is attributed to longer C chains (like C$_8$) trapped between dimer rows. In contrast to that, results of our calculations suggest that nucleation of CVD graphene on a flat Ge(001) surface takes place predominantly on the DV defects.

The vast majority of C$_2$H$_4$ precursor molecules do not give all H atoms away to the substrate. Calculations indicate that the major unpolymerized diffusing species may be, apart from adsorbed H, C$_2$H$_3$ molecules, with small admixture of C$_2$H and C$_2$. This is because most of C$_2$H$_2$ (acetylene) produced during C$_2$H$_4$ decomposition desorb before they decompose further. The surface reactions do not lead to significant production of Ge adatoms. Moreover, although there are Ge adatoms on the surface (due to high temperatures used in these experiments), they are repelled from the edge of a hydrogen-terminated graphene. This is in strong contrast to the MBE process: adsorption of Ge adatoms at edge sites of hydrogen-free graphene is energetically preferred to adsorption at surface steps. 

\begin{figure}
\includegraphics[width=\textwidth]{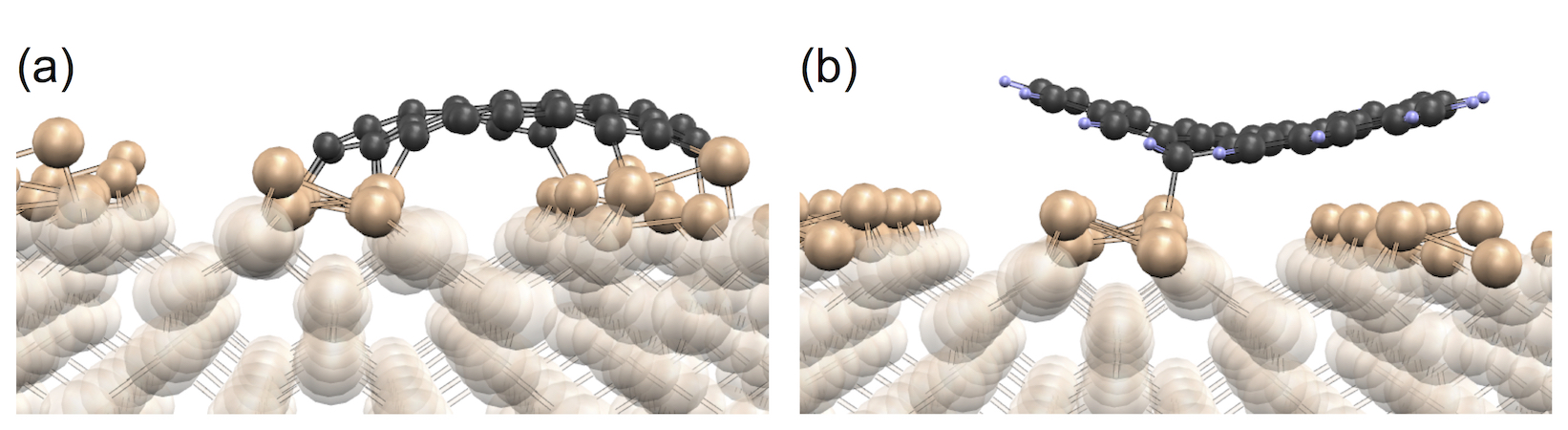}\\
\caption{\label{grmol_Ge} \textbf{Small graphene molecule (29 atoms) on Ge(001)-$p(2\times2)$.} (a) MBE: the edge C atoms tend to make bonds with the substrate. (b) CVD: the edge C atoms are terminated with H, and some of these atoms (here: one atom) makes a weak bond to the substrate.} 
\end{figure}

The C atoms on a clean graphene edge tend to form chemical bonds with the substrate (Fig.~\ref{grmol_Ge}a), immobilizing the molecules consisting of several carbon rings. On the other hand, when the molecule edge is completely saturated with hydrogen (what our calculations predict to happen during our ethylene CVD synthesis), the graphene-substrate bond formation is suppressed, apart from the possibility that some of the edge C atoms may become attached to the substrate by a weak $sp^3$-like bond (Fig.~\ref{grmol_Ge}b). These bonds are however not strong enough to prevent the molecule from diffusing on the surface during the growth process, so that the orientation transfer from the substrate to CVD graphene may occur only during nucleation on immobile seeds, like on the DV defects.

\section*{Discussion}

\begin{figure}
\includegraphics[width=\textwidth]{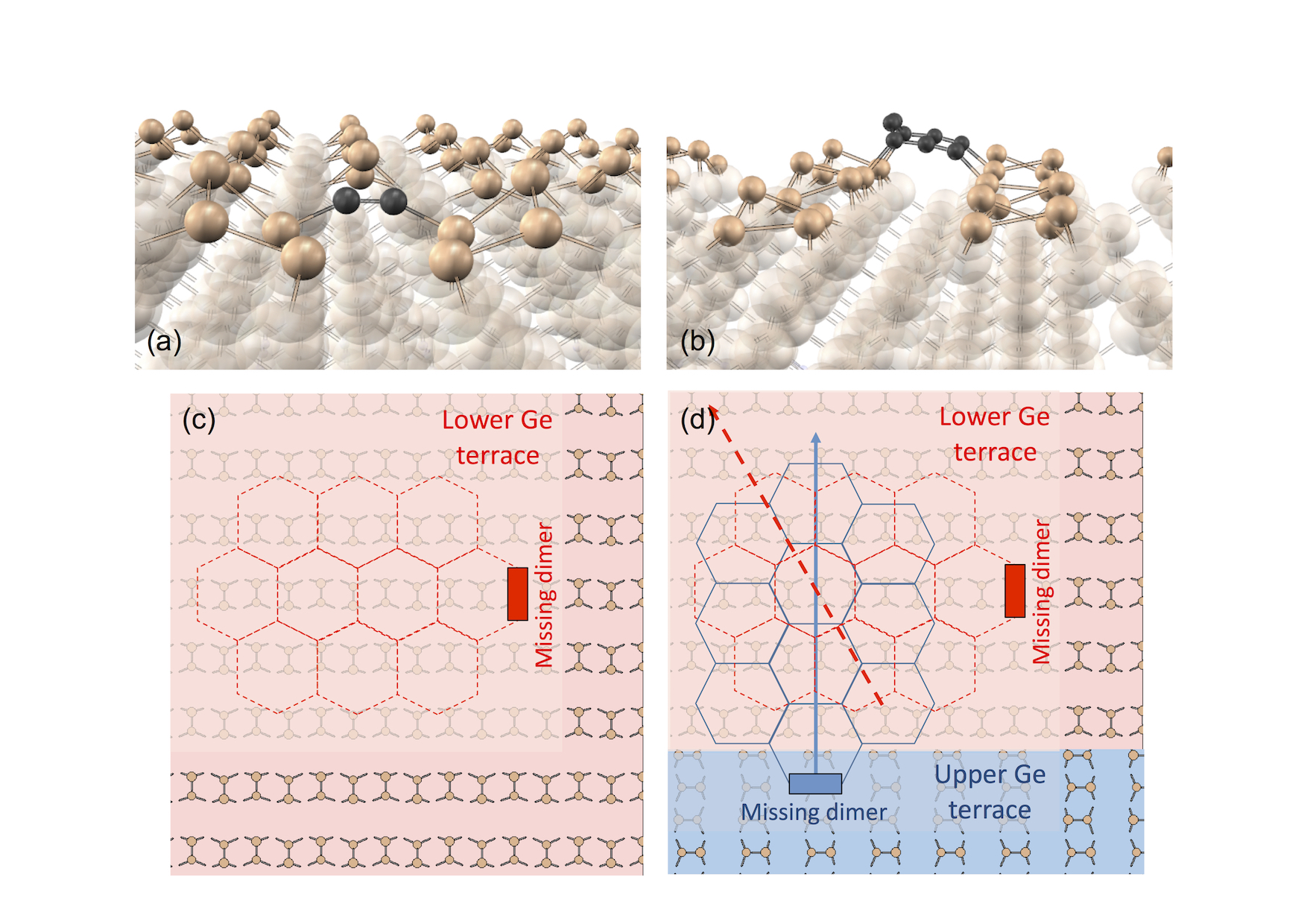}\\
\caption{\label{structure} \textbf{Transfer of crystallographic information from Ge(001)-$p(2\times2)$ to MBE or CVD grown graphene.} (a) C$_2$ dimer, (C$_2$)$_\mathrm{DV}$, trapped in a Ge dimer vacancy. (b) C$_8$ string between dimer rows. (c) Single Ge(001) terrace: Graphene nucleates on a Ge missing dimer (panel a) and continues to grow with the orientation determined by the substrate dimer rows. (d) Two terraces separated by a monatomic step. Graphene sheets originating from both terraces are rotated by $30^\circ$ with respect to one another.} 
\end{figure}

\textbf{Formation mechanism of two domains in graphene growth.} In the MBE process, graphene may nucleate on C dimers, (C$_2$)$_\mathrm{DV}$, trapped inside Ge dimer vacancy defects (Fig.~\ref{structure}a). DV defects are abundant in STM images of Ge(001) even at room-temperature. At $930^\circ$\,C, C$_2$ dimers reside about $50$ times longer in the trapped than in the diffusing state~\cite{Comment_1}. (C$_2$)$_\mathrm{DV}$ is produced also by atomic C in two kick-out events~\cite{Lippert:2014fc,Comment_2}. Dimer vacancies collect carbon; e.\,g., when the second C$_2$ is trapped by a DV already occupied by C$_2$, the energy decreases by $0.9$\,eV, which is similar to the energy gained by trapping of the first C$_2$~\cite{Comment_1}. Trapping of carbon by dimer vacancies provides a means to align the orientation of graphene with that of the substrate (Fig.~\ref{structure}c,d). Variations in the orientation of MBE graphene may be enhanced by nucleation on longer C strings (like C$_8$, Fig.~\ref{structure}b), which prefer to reside between dimer rows and may initiate the growth of graphene in directions somewhat misaligned with dimer rows.

The contribution of Ge DV to graphene nucleation during CVD is less direct, although -- as it will be argued -- important. During the first moments after the exposure of Ge(001) to ethylene, DVs have no influence: at the growth conditions, they do not trap C$_2$H, C$_2$H$_2$, or C$_2$H$_3$. Rings such as C$_6$H$_4$ or C$_{10}$H$_6$ can be produced on the flat surface by polymerization of the direct products of C$_2$H$_4$ decomposition. These arenes stand vertically on the surface and are strongly bonded to the Ge(001) dimers by GeC bonds. But if a polyarene has grown large enough to lean towards the horizontal position as the result of van der Waals attraction, the GeC bonds become weakened and are now vulnerable to attack by H atoms. The latter leave the Ge(001) dimers and stick to C. We estimate that, at the high temperatures needed for the CVD growth, even when the H originates only from C$_2$H$_4$ decomposition and when the partial pressure of C$_2$H$_4$ is only $2\times10^{-2}$\,mbar (as in our experiments), the whole edge of such a polyarene becomes saturated with H (Fig.~\ref{grmol_Ge}b) and the molecule can move freely. It is therefore not obvious how the crystallographic information is transferred from Ge(001) to graphene grown from C$_2$H$_4$, as we observe in experiment (Fig.~\ref{ARPES}). 

\begin{figure}
\includegraphics[width=\textwidth]{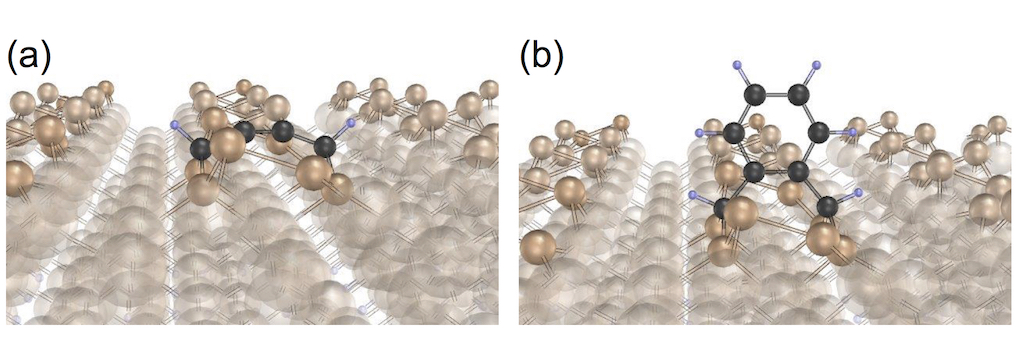}\\
\caption{\label{grmol_DV} \textbf{Influence of a dimer vacancy defect on the graphene growth.} (a) C$_4$H$_2$ and (b) C$_8$H$_6$ molecules inside a dimer vacancy defect on Ge(001)-$p(2\times2)$.} 
\end{figure}

In order to preserve its orientational correlation with the substrate, a ``graphene molecule'' must -- until it grows to a size precluding its diffusion on the surface -- retain the GeC bonds that prevented its small seed from evaporating. This is where the DV defects play an important role. It turns out that DV sites catalyze polymerization of C$_2$H molecules. First, the process of C$_4$H$_2$ formation is significantly more favorable inside a DV (Fig.~\ref{grmol_DV}a) than on the perfect surface: the energy gain increases from $1.6$\,eV to $2.1$\,eV. C$_4$H$_2$ (DV) traps C$_4$H$_4$; the resulting C$_8$H$_6$ molecule is likely to remain inside the dimer vacancy (Fig.~\ref{grmol_DV}b). The GeC bonds are stronger inside a DV than on the perfect surface, and the local geometry of the defect makes them less exposed to H attack. A graphene flake having orientation correlated to that of the orientation of the DV defect may grow in this way. The probability that the nucleus becomes detached before its orientation is fixed by van der Waals forces increases however with the background concentration of H.

\textbf{Crystallographic and electrical quality of graphene.} ARPES measurements strongly indicate that the orientational domains are so small that many of them fit into the area sampled by the X-ray beam spot, i.\,e. into space of about $100\times100\,\mathrm{nm}^2$. The presence of two orientational domains, or more precisely, of grain boundaries between these domains, is thus a factor limiting the electrical quality of graphene grown on Ge(001)/Si(001). According to the mechanism proposed above, the domain size is associated with the abundance of Ge dimer vacancies (on which graphene nucleates), and with the presence of surface steps on the substrate surface (which are responsible for the appearance of graphene with two crystallographic orientations). When the surface is well oriented and flat (local misorientation of $\alpha = 0.5^\circ$), the average step-step distance (or terrace width) equals to $a_s / (\tan(\alpha)\sqrt{2}) \approx 15\,\mathrm{nm}$, where $a_s$ is the surface lattice constant of Ge(001). This is in accordance with the observation that both orientations occur with the same frequency on the area covered by the X-ray spot. The quality of graphene increases with the substrate temperature because the concentration of dimer vacancies $N_\mathrm{DV}$ (thermally activated with $E_\mathrm{form}(\mathrm{DV}) \approx 0.6$\,eV) increases with temperature slower than is the decrease of the fraction $p_\mathrm{occ}(\mathrm{DV},x)$ of DVs occupied with hydrocarbons $x$. For example at thermal equilibrium, $p_\mathrm{occ}(\mathrm{DV},\mathrm{C}_8\mathrm{H}_6)$ is thermally activated with $E_\mathrm{trap}(\mathrm{C}_4\mathrm{H}_2) \approx 1.7$\,eV. In other words, the concentration $s \sim p_\mathrm{occ}N_\mathrm{DV}$ of the growing nuclei decreases with temperature.

\begin{figure}
\includegraphics[width=\textwidth]{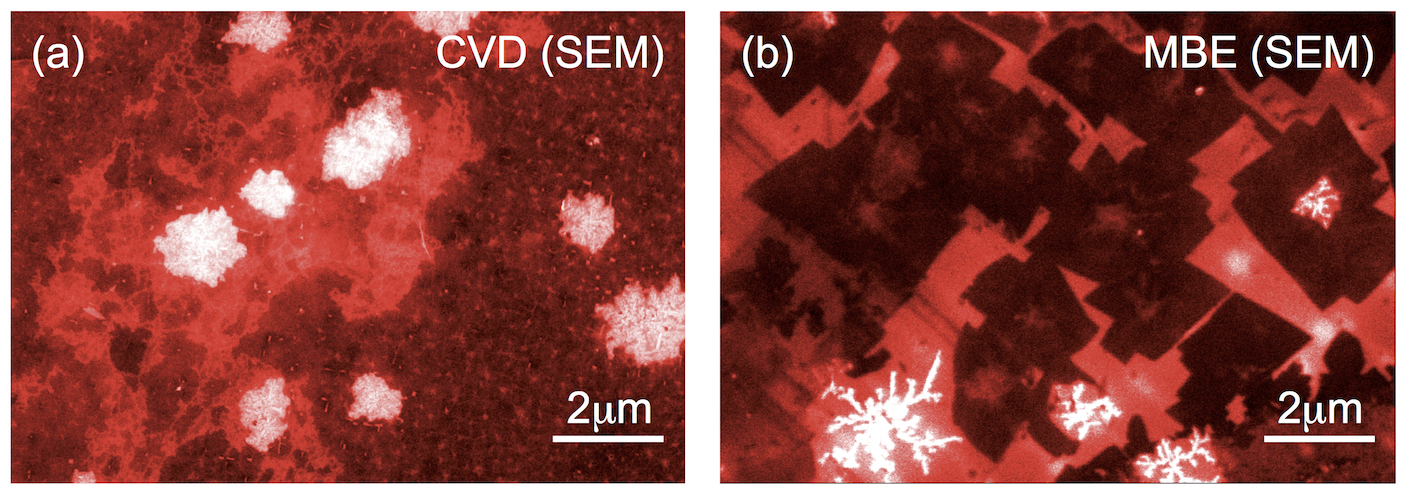}\\
\caption{\label{SEM_maps} \textbf{Distribution of graphene flakes in the CVD and MBE synthesised samples on Ge(001).} Conductive graphene is dark in SEM images for (a) CVD and (b) MBE grown graphene.} 
\end{figure}

Small domain size is responsible also for the curved shape of CVD graphene edges (Fig.~\ref{SEM_maps}). The sharp form of MBE graphene edges (Fig.~\ref{SEM_maps}) is probably an indication that the graphene-like interfacial layer underlying most of the MBE graphene~\cite{Lippert:2014fc} contains much smaller concentration of efficient nucleation seeds, so that the domains may grow much larger, in principle reaching even into the $\mu$m range.

\textbf{Doping and strain.} Doping of the MBE film may originate solely from molecules from the ambient air that are adsorbed on top of the graphene. Higher doping of the CVD film may be caused by germanium atoms~\cite{Emtsev:2011fo}, since the CVD graphene remains in direct contact with Ge(001). 

The compressive strain in the MBE film (-0.3\%) may possibly be caused by thermal expansion coefficient difference between graphene $(-3\times10^{-6}\,\mathrm{K}^{-1})$ and Ge $(6\times10^{-6}\,\mathrm{K}^{-1})$. Cooling from $930^\circ$\,C to $25^\circ$\,C would produce strain of about -0.8\%, which by its absolute value is higher than the -0.3\% estimated under the assumption of biaxial strain (Fig.~\ref{raman}b). It follows that the graphene film forming on the interfacial graphene-like layer grows under tensile strain, possibly induced by the strain in the interfacial layer. Indeed, the strain in the CVD film, which remains in the direct contact with Ge(001), is tensile, indicating that the strain in the growing graphene is also tensile and higher than that the +0.4\% measured at room temperature. This may be caused by coalescence of the small graphene domains.

\section*{Conclusion}

Here we report on the successful growth of graphene on the CMOS compatible Ge(001)/Si(001) substrates from C$_2$H$_4$ precursor in a cold-wall CVD chamber. Experimental results are interpreted on basis of \textit{ab initio} DFT calculations. Presented mechanisms of the graphene synthesis are deduced from experimental and theoretical data.

The major experimental result is that the graphene film on Ge(001)/Si(001) consists of two orientational domains rotated by $30^\circ$ with respect to each other, and that the grain size is comparable to step-step separation of the substrate. Poor crystallographic quality (high Raman D mode peak) and poor electrical quality (low carrier mobility) of the CVD graphene produced in this process are attributed to the small domain size. Arguments linking the grain size with the size of surface terraces separated by Ge(001) monatomic steps are presented. It was argued that the dimer vacancies omnipresent on the surface act as seeds on which graphene can nucleate and that the orientation of a domain is determined by the orientation of Ge dimers on the surface terrace in which the domain nucleated. Small graphene nuclei released from the dimer vacancy sites by hydrogen atoms are volatile; this contributes to the reduction of graphene growth rate by increased surface concentration of hydrogen and to the increase of the average domain size. For completeness and illustration, the CVD growth mode and mechanism was compared to those for the growth from atomic carbon in MBE chamber.

\section*{Methods}

\textbf{Samples preparation.} Graphene was grown in a multi chamber ultrahigh vacuum system (UHV) with a base pressure in the lower $10^{-10}$\,mbar range. The substrate was a Ge(001) film ($2$\,$\mu$m) grown by CVD on a Si(001) wafer and then overgrown \textit{in situ} by epitaxial Ge buffer (about $4$\,nm). The substrate temperature during graphene growth was close to but below the melting point of bulk germanium (it was around $930^\circ$\,C). The substrate was heated from the backside by a graphite filament and the temperature was calibrated by pyrometer. The typical growth rate of the graphene film was in range of a complete monatomic layer within minutes. In addition, graphene samples were grown by a physical method in a MBE chamber (DCA), as described in Ref.~\cite{Lippert:2014fc}: carbon atoms were evaporated via $e$-gun, and the pressure was in the lower $10^{-7}$\,mbar range. The CVD process was done by using ethylene (C$_2$H$_4$) precursor in a UHV-CVD chamber (Riber). The pressure was in the range of $10^{-2}$\,mbar during deposition and the flow of ethylene was $5$\,sccm.

\textbf{Sample characterization.} Quality of graphene was evaluated by $\mu$-Raman spectroscopy using Renishaw In-viaFlex spectrometer and green laser light ($514$\,nm) with a spot diameter of about $500$\,nm. Transport measurements were performed with a 4-tip scanning tunnelling microscope (4-tip STM, Omicron) with a high resolution scanning electron microscope (SEM) used as a tip-navigation tool.

\textbf{ARPES experiments.} Photoemission experiments were performed at the nano-ARPES microscope of the ANTARES beamline of the SOLEIL synchrotron, equipped with two Fresnel zone plates (FZP; i.\,e. for low and high photon energy) for focalization of the synchrotron radiation beam and an order selection aperture (OSA) to eliminate higher diffraction orders. Samples are mounted on a nano-positioning stage placed at the coincident focus of the Scienta R4000 analyser and the FZP focal point, which is responsible for the sample rastering during the image acquisition. The spatial resolution is determined by the FZP resolution and the mechanical stability of the sample stage. Experiments were performed at photon energies $h\nu = 100$\,eV with linearly polarized light for Fermi surface mapping and at $h\nu = 30$\,eV with circularly polarized light for the energy scans around the $\mathrm{K}$ point of graphene. The circular polarization in these experiments was selected in order to avoid the ``dark corridor'' problem along the $\Gamma-\mathrm{K}$ direction for graphene~\cite{Gierz:2011do}. The sample was kept at room temperature during ARPES experiments. The base pressure during all experiments was better than $5\times10^{-11}$\,mbar.

\textbf{DFT calculations.} Pseudopotential density functional theory (DFT) total energies and atomic and electronic structures were calculated by Quantum Espresso~\cite{Giannozzi:2009hx}. Exchange and correlation energy was approximated by the PBE gradient-corrected functional~\cite{Perdew:1996}. Periodically repeated slab consisting of eight Ge(001) planes separated by $\approx10\,\mathrm{\AA}$ of the empty space was used. The calculations were done for square Ge(001)-$p(2\times2)$ surface cells, each containing eight dimers on one side and passivated with H atoms on the other side. Further calculation details can be found elsewhere~\cite{Lippert:2014fc}.

%\section*{References}

%\bibliography{/Users/YuDedkov/Work/Articles/___REFERENCES___/references_all.bib}

\begin{thebibliography}{10}
\expandafter\ifx\csname url\endcsname\relax
  \def\url#1{\texttt{#1}}\fi
\expandafter\ifx\csname urlprefix\endcsname\relax\def\urlprefix{URL }\fi
\providecommand{\bibinfo}[2]{#2}
\providecommand{\eprint}[2][]{\url{#2}}

\bibitem{Geim:2007a}
\bibinfo{author}{Geim, A.~K.} \& \bibinfo{author}{Novoselov, K.~S.}
\newblock \bibinfo{title}{{The rise of graphene}}.
\newblock \emph{\bibinfo{journal}{Nature Mater.}}
  \textbf{\bibinfo{volume}{6}}, \bibinfo{pages}{183--191}
  (\bibinfo{year}{2007}).

\bibitem{Geim:2009}
\bibinfo{author}{Geim, A.}
\newblock \bibinfo{title}{{Graphene: Status and prospects}}.
\newblock \emph{\bibinfo{journal}{Science}} \textbf{\bibinfo{volume}{324}},
  \bibinfo{pages}{1530--1534} (\bibinfo{year}{2009}).

\bibitem{Bae:2010}
\bibinfo{author}{Bae, S.} \emph{et~al.}
\newblock \bibinfo{title}{{Roll-to-roll production of 30-inch graphene films
  for transparent electrodes}}.
\newblock \emph{\bibinfo{journal}{Nature Nanotech.}}
  \textbf{\bibinfo{volume}{5}}, \bibinfo{pages}{574--578}
  (\bibinfo{year}{2010}).

\bibitem{Bonaccorso:2013iy}
\bibinfo{author}{Bonaccorso, F.} \emph{et~al.}
\newblock \bibinfo{title}{{Production and processing of graphene and 2d
  crystals}}.
\newblock \emph{\bibinfo{journal}{Mater. Today}}
  \textbf{\bibinfo{volume}{15}}, \bibinfo{pages}{564--589}
  (\bibinfo{year}{2013}).

\bibitem{Novoselov:2013hw}
\bibinfo{author}{Novoselov, K.~S.} \emph{et~al.}
\newblock \bibinfo{title}{{A roadmap for graphene}}.
\newblock \emph{\bibinfo{journal}{Nature}} \textbf{\bibinfo{volume}{490}},
  \bibinfo{pages}{192--200} (\bibinfo{year}{2013}).

\bibitem{Yu:2008}
\bibinfo{author}{Yu, Q.} \emph{et~al.}
\newblock \bibinfo{title}{{Graphene segregated on Ni surfaces and transferred
  to insulators}}.
\newblock \emph{\bibinfo{journal}{Appl. Phys. Lett.}}
  \textbf{\bibinfo{volume}{93}}, \bibinfo{pages}{113103}
  (\bibinfo{year}{2008}).

\bibitem{Kim:2009a}
\bibinfo{author}{Kim, K.~S.} \emph{et~al.}
\newblock \bibinfo{title}{{Large-scale pattern growth of graphene films for
  stretchable transparent electrodes}}.
\newblock \emph{\bibinfo{journal}{Nature}} \textbf{\bibinfo{volume}{457}},
  \bibinfo{pages}{706--710} (\bibinfo{year}{2009}).

\bibitem{Li:2009}
\bibinfo{author}{Li, X.} \emph{et~al.}
\newblock \bibinfo{title}{{Large-area synthesis of high-quality and uniform
  graphene films on copper foils}}.
\newblock \emph{\bibinfo{journal}{Science}} \textbf{\bibinfo{volume}{324}},
  \bibinfo{pages}{1312--1314} (\bibinfo{year}{2009}).

\bibitem{Ambrosi:2014gl}
\bibinfo{author}{Ambrosi, A.} \& \bibinfo{author}{Pumera, M.}
\newblock \bibinfo{title}{{The CVD graphene transfer procedure introduces
  metallic impurities which alter the graphene electrochemical properties}}.
\newblock \emph{\bibinfo{journal}{Nanoscale}} \textbf{\bibinfo{volume}{6}},
  \bibinfo{pages}{472--476} (\bibinfo{year}{2014}).

\bibitem{Lupina:2015je}
\bibinfo{author}{Lupina, G.} \emph{et~al.}
\newblock \bibinfo{title}{{Residual metallic contamination of transferred
  chemical vapor deposited graphene}}.
\newblock \emph{\bibinfo{journal}{ACS Nano}} \textbf{\bibinfo{volume}{9}},
  \bibinfo{pages}{4776--4785} (\bibinfo{year}{2015}).

\bibitem{Ochedowski:2012bx}
\bibinfo{author}{Ochedowski, O.} \emph{et~al.}
\newblock \bibinfo{title}{{Graphene on Si(111) $7\times7$}}.
\newblock \emph{\bibinfo{journal}{Nanotechnology}}
  \textbf{\bibinfo{volume}{23}}, \bibinfo{pages}{405708}
  (\bibinfo{year}{2012}).

\bibitem{ThanhTrung:2014hd}
\bibinfo{author}{Trung, P.~T.} \emph{et~al.}
\newblock \bibinfo{title}{{Direct growth of graphene on Si(111)}}.
\newblock \emph{\bibinfo{journal}{J. Appl. Phys.}}
  \textbf{\bibinfo{volume}{115}}, \bibinfo{pages}{223704}
  (\bibinfo{year}{2014}).

\bibitem{Dang:2015fk}
\bibinfo{author}{Dang, X.} \emph{et~al.}
\newblock \bibinfo{title}{{Semiconducting graphene on silicon from
  first-principles calculations}}.
\newblock \emph{\bibinfo{journal}{ACS Nano}} \textbf{\bibinfo{volume}{9}},
  \bibinfo{pages}{8562--8568} (\bibinfo{year}{2015}).

\bibitem{Zeng:2013jp}
\bibinfo{author}{Zeng, L.-H.} \emph{et~al.}
\newblock \bibinfo{title}{{Monolayer graphene/germanium Schottky junction as
  high-performance self-driven infrared light photodetector}}.
\newblock \emph{\bibinfo{journal}{ACS Appl. Mater. Interfaces}}
  \textbf{\bibinfo{volume}{5}}, \bibinfo{pages}{9362--9366}
  (\bibinfo{year}{2013}).

\bibitem{Cavallo:2014fp}
\bibinfo{author}{Cavallo, F.} \emph{et~al.}
\newblock \bibinfo{title}{{Exceptional charge transport properties of graphene
  on germanium}}.
\newblock \emph{\bibinfo{journal}{ACS Nano}} \textbf{\bibinfo{volume}{8}},
  \bibinfo{pages}{10237--10245} (\bibinfo{year}{2014}).

\bibitem{Wang:2013fq}
\bibinfo{author}{Wang, G.} \emph{et~al.}
\newblock \bibinfo{title}{{Direct growth of graphene film on germanium
  substrate}}.
\newblock \emph{\bibinfo{journal}{Sci. Rep.}} \textbf{\bibinfo{volume}{3}},
  \bibinfo{pages}{2465} (\bibinfo{year}{2013}).

\bibitem{Lee:2014dv}
\bibinfo{author}{Lee, J.-H.} \emph{et~al.}
\newblock \bibinfo{title}{{Wafer-scale growth of single-crystal monolayer
  graphene on reusable hydrogen-terminated germanium.}}
\newblock \emph{\bibinfo{journal}{Science}} \textbf{\bibinfo{volume}{344}},
  \bibinfo{pages}{286--289} (\bibinfo{year}{2014}).

\bibitem{Jacobberger:2015de}
\bibinfo{author}{Jacobberger, R.~M.} \emph{et~al.}
\newblock \bibinfo{title}{{Direct oriented growth of armchair
  graphenenanoribbons on germanium}}.
\newblock \emph{\bibinfo{journal}{Nature Commun.}}
  \textbf{\bibinfo{volume}{6}}, \bibinfo{pages}{8006} (\bibinfo{year}{2015}).

\bibitem{Lippert:2014fc}
\bibinfo{author}{Lippert, G.} \emph{et~al.}
\newblock \bibinfo{title}{{Graphene grown on Ge(001) from atomic source}}.
\newblock \emph{\bibinfo{journal}{Carbon}} \textbf{\bibinfo{volume}{75}},
  \bibinfo{pages}{104--112} (\bibinfo{year}{2014}).

\bibitem{McElhinny:2016gw}
\bibinfo{author}{McElhinny, K.~M.}, \bibinfo{author}{Jacobberger, R.~M.},
  \bibinfo{author}{Zaug, A.~J.}, \bibinfo{author}{Arnold, M.~S.} \&
  \bibinfo{author}{Evans, P.~G.}
\newblock \bibinfo{title}{{Graphene-induced Ge (001) surface faceting}}.
\newblock \emph{\bibinfo{journal}{Surf. Sci.}} (\bibinfo{year}{2016}); \bibinfo{pages}{doi: 10.1016/j.susc.2015.12.035}.
  
\bibitem{Pasternak:2016jt}
\bibinfo{author}{Pasternak, I.} \emph{et~al.}
\newblock \bibinfo{title}{{Graphene growth on Ge(100)/Si(100) substrates by CVD method}}.
\newblock \emph{\bibinfo{journal}{Sci. Rep.}} \textbf{\bibinfo{volume}{6}}, \bibinfo{pages}{21773}
  (\bibinfo{year}{2016}).

\bibitem{Yang:2013eva}
\bibinfo{author}{Yang, W.} \emph{et~al.}
\newblock \bibinfo{title}{{Epitaxial growth of single-domain graphene on
  hexagonal boron nitride}}.
\newblock \emph{\bibinfo{journal}{Nature Mater.}}
  \textbf{\bibinfo{volume}{12}}, \bibinfo{pages}{792--797}
  (\bibinfo{year}{2013}).

\bibitem{Mehr:2012cv}
\bibinfo{author}{Mehr, W.} \emph{et~al.}
\newblock \bibinfo{title}{{Vertical graphene base transistor}}.
\newblock \emph{\bibinfo{journal}{IEEE Electr. Device Lett.}}
  \textbf{\bibinfo{volume}{33}}, \bibinfo{pages}{691--693}
  (\bibinfo{year}{2012}).

\bibitem{Vaziri:2013er}
\bibinfo{author}{Vaziri, S.} \emph{et~al.}
\newblock \bibinfo{title}{{A graphene-based hot electron transistor}}.
\newblock \emph{\bibinfo{journal}{Nano Lett.}} \textbf{\bibinfo{volume}{13}},
  \bibinfo{pages}{1435--1439} (\bibinfo{year}{2013}).

\bibitem{Avila:2013a}
\bibinfo{author}{Avila, J.} \emph{et~al.}
\newblock \bibinfo{title}{{Interferometer-controlled soft X-ray scanning photoemission microscope at SOLEIL}}.
\newblock \emph{\bibinfo{journal}{J. Phys.: Conf. Ser.}} \textbf{\bibinfo{volume}{425}},
  \bibinfo{pages}{132013} (\bibinfo{year}{2013}).

\bibitem{Avila:2013b}
\bibinfo{author}{Avila, J.} \emph{et~al.}
\newblock \bibinfo{title}{{ANTARES, a scanning photoemission microscopy beamline at SOLEIL.}}.
\newblock \emph{\bibinfo{journal}{J. Phys.: Conf. Ser.}} \textbf{\bibinfo{volume}{425}},
  \bibinfo{pages}{192023} (\bibinfo{year}{2013}).

\bibitem{Avila:2014a}
\bibinfo{author}{Avila, J.} \& \bibinfo{author}{Asensio, M.~C.}
\newblock \bibinfo{title}{{First nanoARPES user facility available at SOLEIL: An innovative and powerful tool for studying advanced materials}}.
\newblock \emph{\bibinfo{journal}{Synchr. Rad. News}} \textbf{\bibinfo{volume}{27}},
  \bibinfo{pages}{24--30} (\bibinfo{year}{2014}).

\bibitem{Emtsev:2011fo}
\bibinfo{author}{Emtsev, K.~V.}, \bibinfo{author}{Zakharov, A.~A.},
  \bibinfo{author}{Coletti, C.}, \bibinfo{author}{Forti, S.} \&
  \bibinfo{author}{Starke, U.}
\newblock \bibinfo{title}{{Ambipolar doping in quasifree epitaxial graphene on
  SiC(0001) controlled by Ge intercalation}}.
\newblock \emph{\bibinfo{journal}{Phys. Rev. B}} \textbf{\bibinfo{volume}{84}},
  \bibinfo{pages}{125423} (\bibinfo{year}{2011}).

\bibitem{Lee:2012gy}
\bibinfo{author}{Lee, J.~E.}, \bibinfo{author}{Ahn, G.}, \bibinfo{author}{Shim,
  J.}, \bibinfo{author}{Lee, Y.~S.} \& \bibinfo{author}{Ryu, S.}
\newblock \bibinfo{title}{{Optical separation of mechanical strain from charge
  doping in graphene}}.
\newblock \emph{\bibinfo{journal}{Nature Commun.}}
  \textbf{\bibinfo{volume}{3}}, \bibinfo{pages}{1024}
  (\bibinfo{year}{2012}).

\bibitem{Frank:2011cu}
\bibinfo{author}{Frank, O.} \emph{et~al.}
\newblock \bibinfo{title}{{Raman 2D-band splitting in graphene: Theory and
  experiment}}.
\newblock \emph{\bibinfo{journal}{ACS Nano}} \textbf{\bibinfo{volume}{5}},
  \bibinfo{pages}{2231--2239} (\bibinfo{year}{2011}).

\bibitem{Pisana:2007cl}
\bibinfo{author}{Pisana, S.} \emph{et~al.}
\newblock \bibinfo{title}{{Breakdown of the adiabatic
  Born{\textendash}Oppenheimer approximation in graphene}}.
\newblock \emph{\bibinfo{journal}{Nature Mater.}}
  \textbf{\bibinfo{volume}{6}}, \bibinfo{pages}{198--201}
  (\bibinfo{year}{2007}).

\bibitem{Lazzeri:2006kz}
\bibinfo{author}{Lazzeri, M.} \& \bibinfo{author}{Mauri, F.}
\newblock \bibinfo{title}{{Nonadiabatic Kohn anomaly in a doped graphene
  monolayer}}.
\newblock \emph{\bibinfo{journal}{Phys. Rev. Lett.}}
  \textbf{\bibinfo{volume}{97}}, \bibinfo{pages}{266407}
  (\bibinfo{year}{2006}).

\bibitem{Comment_1}
Assuming thermal equilibrium DV concentration at $900^\circ$\,C, the DV formation energy of $0.6$\,eV, and the same attempt rates for C$_2$ diffusion and C$_2$ escape from DV. In order to escape from the trap, a C$_2$ must overcome a barrier of $2.3$\,eV, higher by $1.0$\,eV than the barrier for C$_2$ diffusion on the perfect surface. The total energy of C$_2$ is by $1.0$\,eV lower in a dimer vacancy at the most favorable adsorption site on the perfect surface.

\bibitem{Comment_2}
C(diff) $\rightarrow$  C$_\mathrm{Ge}$ + Ge(bulk) - $0.13$\,eV, and C(diff) + C$_\mathrm{Ge}$ $\rightarrow$ (C$_2$)$_\mathrm{DV}$ + Ge(bulk)  + $0.31$\,eV, where C(diff) is the mobile atomic carbon diffusing under the surface and Ge(bulk) is a Ge atom having the same energy as in the Ge bulk. The total energy gain of $0.2$\,eV is in this case much smaller than for trapping on a DV ($1.0$\,eV~\cite{Comment_1}), because a new DV site must be produced, by subsequent ejection of two Ge atoms. The difference between these two energies ($0.8$\,eV) should thus be equal to the DV formation energy ($0.6$\,eV~\cite{Comment_1}). The discrepancy ($0.2$\,eV) comes from inaccuracies in the calculation of atomic relaxations, caused by the presence of nearly degenerate metastable configurations of defects and by finite size effects. This inaccuracy amounts to at least $0.1$\,eV per defect.

\bibitem{Gierz:2011do}
\bibinfo{author}{Gierz, I.}, \bibinfo{author}{Henk, J.},
  \bibinfo{author}{H{\"o}chst, H.}, \bibinfo{author}{Ast, C.~R.} \&
  \bibinfo{author}{Kern, K.}
\newblock \bibinfo{title}{{Illuminating the dark corridor in graphene:
  Polarization dependence of angle-resolved photoemission spectroscopy on
  graphene}}.
\newblock \emph{\bibinfo{journal}{Phys. Rev. B}} \textbf{\bibinfo{volume}{83}},
  \bibinfo{pages}{121408} (\bibinfo{year}{2011}).
  
\bibitem{Giannozzi:2009hx}
\bibinfo{author}{Giannozzi, P.} \emph{et~al.}
\newblock \bibinfo{title}{{QUANTUM ESPRESSO: a modular and open-source software project for quantum simulations of materials}}.
\newblock \emph{\bibinfo{journal}{J. Phys.: Condens. Matter}}
  \textbf{\bibinfo{volume}{21}}, \bibinfo{pages}{395502}
  (\bibinfo{year}{2009}).

\bibitem{Perdew:1996}
\bibinfo{author}{Perdew, J.~P.}, \bibinfo{author}{Burke, K.} \&
  \bibinfo{author}{Ernzerhof, M.}
\newblock \bibinfo{title}{{Generalized gradient approximation made simple}}.
\newblock \emph{\bibinfo{journal}{Phys. Rev. Lett.}} \textbf{\bibinfo{volume}{77}},
  \bibinfo{pages}{3865--3868} (\bibinfo{year}{1996}).

\end{thebibliography}

\section*{Acknowledgements}

The authors wish to dedicate this work to the memory of Prof. Wolfgang Mehr, the inventor and the pioneer of research on graphene base transistors. This work was supported by the European Commission through the GRADE project (No. 317839) and via ERC grant no. 259286. Computing time support from the J\"ulich Supercomputing Center of the John von Neumann Institute for Computing (project hfo06) is gratefully acknowledged. 

\section*{Author contributions statement}

J.D. performed DFT modeling. G.L., G.L., T.S., M.S., D.V. prepared graphene on Ge substrates by CVD and MBE methods. G.L., J.A., I.C., M.S., D.V., M.-C.A. performed nano-ARPES measurements. J.B., C.T. carried out 4-tip STM characterizations. F.H., J.M. performed Raman experiments. J.D., G.L., J.A., J.B., I.C., Yu.S.D., F.H., G.L., J.M., T.S., T.S., M.S., C.T., D.V., M.-C.A. analyzed the data and contribute in the preparation of the manuscript.

\section*{Additional information}

\textbf{Competing financial interests:} The authors declare no competing financial interests.

\clearpage

\clearpage

\clearpage

\clearpage

\clearpage

\clearpage

\clearpage

\clearpage

%%%

\end{document}